# Experimental identification of critical condition for drastically enhancing thermoelectric power factor of two-dimensional layered materials

Junwen Zeng,[†,⊥,#] Xin He,[‡,#] Shi-Jun Liang,[*,†,#] Erfu Liu,[†] Yuanhui Sun,[‡] Chen Pan,[†] Yu Wang,[†] Tianjun Cao,[†] Xiaowei Liu,[†] Chengyu Wang,[†] Lili Zhang,[†] Shengnan Yan,[†] Guangxu Su,[†] Zhenlin Wang,[†] Kenji Watanabe,[§] Takashi Taniguchi,[§] David J. Singh[∥,‡], Lijun Zhang[*,‡] and Feng Miao[*,†]

[†] National Laboratory of Solid State Microstructures, School of Physics, Collaborative Innovation Center of Advanced Microstructures, Nanjing University, Nanjing 210093, China

[‡] Key Laboratory of Automobile Materials of MOE, State Key Laboratory of Superhard Materials, and School of Materials Science, Jilin University, Changchun 130012, China

[§] National Institute for Materials Science, 1-1 Namiki Tsukuba, Ibaraki 305-0044, Japan

[∥] Department of Physics and Astronomy, University of Missouri, Columbia, MO 65211-7010 USA

[⊥] Center for Excellence in Superconducting Electronics, State Key Laboratory of Functional Material for Informatics, Shanghai Institute of Microsystem and Information Technology, Chinese Academy of Sciences, Shanghai 200050, China

[#] These authors contributed equally to this work.

[*] E-mail: miao@nju.edu.cn (F.M.).

[*] E-mail: lijun_zhang@jlu.edu.cn (L.Z.).

[*] E-mail: sjliang@nju.edu.cn (S.J.L.).

**ABSTRACT:** Nano-structuring is an extremely promising path to high performance thermoelectrics. Favorable improvements in thermal conductivity are attainable in many material systems, and theoretical work points to large improvements in electronic properties. However, realization of the electronic benefits in practical materials has been elusive experimentally. A key challenge is that experimental identification of the quantum confinement length, below which the thermoelectric power factor is significantly enhanced, remains elusive due to lack of simultaneous control of size and carrier density. Here we investigate gate-tunable and temperature-dependent thermoelectric transport in γ-phase indium selenide (γ-InSe, n-type semiconductor) samples with thickness varying from 7 to 29 nm. This allows us to properly map out dimension and doping space. Combining theoretical and experimental studies, we reveal that the sharper pre-edge of the conduction-band density of states arising from quantum confinement gives rise to an enhancement of the Seebeck coefficient and the power factor in the thinner InSe samples. Most importantly, we experimentally identify the role of the competition between quantum confinement length and thermal de Broglie wavelength in the enhancement of power factor. Our results provide an important and general experimental guideline for optimizing the power factor and improving the thermoelectric performance of two-dimensional layered semiconductors.

The performance of thermoelectric materials is determined by a figure of merit, $ZT = S^2\sigma T/\kappa$, where $S$ is the Seebeck coefficient, which gives the open circuit voltage per unit temperature difference, $\sigma$ is the electrical conductivity, $T$ is temperature, and $\kappa$ denotes the thermal conductivity[1-3]. Optimization of electrical transport properties $S^2\sigma$, also known as the power factor (PF), and reduction of lattice thermal conductivity are effective ways to achieve high $ZT$[4-7]. Maximizing $S^2\sigma$ relies on engineering electronic band structures of materials, optimization of carrier concentration, and manipulation of carrier scattering. Much attention has been paid to the engineering of the band structure. Importantly, it was proposed initially by Hicks and Dresselhaus that the enhanced electronic density of states (DOS) per unit volume in two-dimensional (2D) quantum well and superlattice materials would give rise to highly increased power factors over the bulk values[7,8]. However, it has been a big challenge to experimentally realize large increases in *ZT* over normal high-performance bulk thermoelectrics by this strategy. Some previous experiments, in which the quantum well width in the crystalline semiconductor is comparable to the charge carrier Bohr radius, showed that the power factor is not noticeably enhanced with reducing dimension of samples[9,10], leading to a scenario on whether 2D quantum confinement is a significant approach to enhancing power factor. Recent theory argued that the discrepancy between theory and experiments arises from a competition between the quantum confinement length $L$ and the thermal de Broglie wavelength $\xi$ [11,12]. Furthermore, it should be noted that enhanced DOS, while favorable for achieving high *S*, may also increase the carrier scattering, complicating the analysis. For thermoelectric materials with 2D quantum confinement, the power factor can only be considerably enhanced when the ratio of $L/\xi$ is far less than 1. Moreover, many thermoelectric materials have complex band structures that are not well approximated by simple parabolic band models, again complicating the analysis. Therefore, it is imperative to have experimental data on carrier concentration dependent transport properties under controlled conditions. However, well-controlled $L$ and gate-tunable carrier density are inaccessible in the conventional quantum well thermoelectric materials, which increases the difficulty in achieving an optimal $L/\xi$ ratio. Compared to the conventional quantum well thermoelectric materials, 2D layered semiconductor materials held together by weak van der Waals interactions can be readily fabricated into high-quality samples with controllable and continuously-varying thickness via top-down mechanical exfoliation[13] or bottom-up chemical vapor deposition approach[14]. The carrier concentrations can be gate-tunable in such 2D layered semiconductors. Therefore, they offer a promising platform to investigate the effects of quantum confinement on enhancement of $S^2\sigma$ and thus thermoelectric performance.

In fact, 2D layered semiconductor materials such as $MoS_2$ and black phosphorus have shown huge potential for high-performance thermoelectric devices[15-23]. 2D Indium selenide has recently been identified as one of high-mobility 2D materials at room temperature[24,25], and its bulk compounds have a large Seebeck coefficient of up to -400 μV/K[26-28]. According to the recent theory[11,12], realization of $L/\xi \leq 1$ may make 2D indium selenide a highly promising thermoelectric material[29], especially in light of its low thermal conductivity[30,31]. Therefore, it is of crucially important to experimentally identify the critical condition associated with thickness of 2D indium selenide (InSe) below which the power factor could be greatly enhanced over its bulk form.

Here we present a joint theory-experiment study of the thickness dependence of thermoelectric transport properties in 2D InSe at various temperatures and gate-tunable carrier densities. We observe that the Seebeck coefficient and the power factor in thinner InSe samples at room temperature are much higher than the reported values of bulk samples. Based on *ab initio* calculations and Boltzmann transport theory, we attribute the enhancement of |*S*| and PF in thinner films to the shaper edge of the conduction-band DOS onset due to quantum confinement. More importantly, we experimentally verify that the power factor in 2D InSe can be drastically enhanced when the confinement length of sample is shorter than the thermal de Broglie wavelength, in agreement with recently-proposed theory[11,12]. Our work offers a roadmap for improving the power factor of 2D semiconductors, and holds promise of paving the way towards realization of high-performance 2D thermoelectric devices. It finally realizes the promise of the initial studies by Hicks and Dresselhaus, and proves their predictions for strongly enhanced power factors in dimensionally reduced superlattices and quantum well thermoelectrics.

γ-phase InSe is a member of layered metal chalcogenide semiconductors with interlayer distance of about 0.8 nm (Figure 1a)[24]. The thickness-dependent Raman spectra are shown in Figure 1b, with four dominant characteristic Raman peaks: $A_1'$, $E'(TO)$, $E'(LO)$ and $A_1$ modes[32]. Reducing the thickness of InSe leads to enhanced Raman intensities. The anomaly of thickness-dependent Raman intensity (from 7 nm to 29 nm) may be attributed to the multi-layer interference (reflection) of excitation (Raman signals)[33] or thickness dependence of the InSe electronic structures[34]. To fabricate high-quality InSe thermoelectric devices, we chose hexagonal boron nitride (hBN) to replace conventional $SiO_2$ as the substrate to suppress the substrate phonon scattering and strong Coulomb scattering arising from charged impurities. To measure $S$, we fabricated an on-chip set-up integrating a micro-fabricated heater and two micro-thermometers. The schematic of device setup and its optical microscope image are shown in Figure 1c-d, respectively. A current is applied to micro-fabricated heater to generate the temperature gradient $\Delta T$ along the InSe channel. The temperature gradient induced voltage $\Delta V$ is measured by the voltage probes between the source and drain electrodes, which also serves as the micro-thermometers. More detailed measurement procedures are shown in Figure S1.

The thermoelectric power factor of InSe is strongly dependent on electrical conductivity. The carrier density-dependence of electrical conductivity in InSe samples can be manifested in transfer characteristic curves (Figure 2a-b), exhibiting a typical n-type electron conduction in InSe. The two-terminal measurements of electrical transport properties in InSe can be greatly affected by the contact properties (Figure 2a). Therefore, we carried out four-terminal resistance $R_{xx}$ measurements via an AC lock-in technique, with the corresponding four-terminal conductivity $\sigma_{xx}$ shown in Figure 2b ($\sigma_{xx} = (L/W)(1/R_{xx})$, $L$, $W$ is the length and width of the sample, respectively). The metallic state is greatly affected by scattering processes in InSe over the entire temperature (90 K - 300 K) and gate voltage (25 V - 50 V) ranges. To understand the specific scattering at play, we extracted the field-effect mobility $\mu_{FE}$ through the

Drude model, $\mu_{FE} = \partial(\sigma_{xx})/\partial(C_g V_{bg})$, where $C_g$ is the capacitance per unit area between the sample and the back gate. The electron mobility monotonically increases with decreasing $T$. It is up to 470 cm² V⁻¹ s⁻¹ at room temperature and increases to 3,560 cm² V⁻¹ s⁻¹ at $T$= 90 K and $n$ = 2.1×10¹² cm⁻² (Inset of Figure 2b). The temperature dependence of mobility follows the conventional $T^{-\gamma}$ relation, with γ ≥ 1 at $n$ =1.8, 2.1×10¹² cm⁻² over the whole temperature range, indicating that the acoustic phonon scattering plays a dominant role for electron transport with possibly non-negligible additional contributions from optical phonons[24,35]. The characteristic of phonon-limited scattering is usually seen in most high performance thermoelectrics.

The scattering mechanism limiting mobility not only affects the electrical conductivity but also the Seebeck coefficient in InSe samples[20,23,36]. Optimization of power factor has been challenging due to the countering variation of conductivity and Seebeck coefficient as a function of doping concentration. This underscores the critical importance of optimizing doping in thermoelectric materials, which is extremely challenging in nanoscale systems by chemical doping. To that end, we examined the carrier density-dependence of Seebeck coefficient in a typical 10 nm thick InSe sample at different temperatures (Figure 2c). The negative $S$ is indicative of the electron-dominant thermoelectric behavior in n-type semiconductors, consistent with the electron conduction demonstrated above (Figure 2a). For a given temperature, we observe that the |$S$| increases with electron density decreasing and that the maximum |$S$| is up to 480 μV/K at room temperature. When the electron density (or back gate voltage) is further reduced down to the subthreshold regime, measurement of $S$ is challenging in the high impedance channel due to high noise level[37]. Decreasing temperature leads to a reduction in |$S$| as typical outside regimes with bipolar conduction. To understand the thermoelectric behavior, we performed first-principles calculations of the Seebeck coefficient $S$ within the framework of the Boltzmann transport theory[38-40] for 12 layers (L) (close to 10 nm thick InSe sample used in the experiment) InSe film (Methods). Note that the temperature-dependent electron mobility $\mu$ (The Inset of Figure 2b) indicates that the acoustic phonon scattering is the dominant electron scattering mechanism at $T \geq 90$ K[24]. The acoustic phonon scattering in the 2D material systems gives rise to the energy ($\varepsilon$) dependent relaxation time (*i.e.,* $\tau(\varepsilon) \propto \varepsilon^r$ with $r = 0$)[19,35], which reflects the flat DOS above the band edge for a 2D parabolic band. The calculated *S* values (dashed lines in Figure 2c) are in quantitatively good agreement with the experimental data (symbols), especially in the region of larger electron concentrations and higher temperatures. The slight deviation at small electron concentrations and low temperatures may originate from other electron scattering mechanisms neglected in the calculations, which must be important at low carrier concentrations, since the different energy dependent scattering rates for other mechanisms, e.g. due to point defects, make them more important at low carrier densities.

As mentioned, engineering band structures to enhance power factors is a promising but challenging avenue for high performance thermoelectrics[7,8,41,42]. Increasingly strong quantum confinement in the 2D InSe through reducing its thickness offers a vital approach to enhance the DOS. We measured room-temperature Seebeck coefficient as a function of gate-tunable electron density (Figure 2d) for samples with different thickness. We observe that the Seebeck coefficient increases with reducing the thickness of sample from 29 nm to 7 nm. Compared with $|S| = 288$ μV/K in the 29 nm thick sample, the magnitude of Seebeck coefficient in the 7 nm thick sample is enhanced to 570 μV/K at the same carrier density of $0.7 \times 10^{12}$ cm$^{-2}$, which is much higher than the reported value (up to 400 μV/K) of bulk indium selenide[26,27]. Note that this is in contrast to the expectation from the bulk carrier concentration, since reducing thickness at a fixed 2D carrier concentration implies an increased effective 3D carrier concentration by assuming a bulk-like sample. The Seebeck coefficients at room temperature are also much higher than that of black phosphorus thin films (up to 400 μV/K) with similar carrier density[17]. To interpret the variation of the measured Seebeck coefficient with the thickness, we calculated the Seebeck coefficients at corresponding thicknesses by combining first principles and Boltzmann transport equation, and then compared theoretical results (dashed lines in Figure 2d) with the experimental data. Notably, our calculations well reproduce the experimental results that reducing thickness of samples enhances the |*S*|. With the good agreement between the theoretical

calculations and the experimental data, it is reasonable to attribute the enhancement of |S| in the thinner InSe films to the increased sharpness of the electronic DOS onset arising from quantum confinement[7,8].

To further prove the correlation between enhancement of |S| and confinement-induced increased DOS, we calculated and compared the electronic structures of the 9L and 36L thick InSe samples (Figure 3a), corresponding to our 7 nm and 29 nm thick samples used in the experiments, respectively. We find that the width of bands, for instance the first group of conduction bands at Γ point, is substantially reduced in the 7 nm thick sample, and that the conduction bands become much less dispersive, which is justified by the increased electron effective mass from 0.145 $m_0$ at 36L to 0.153 $m_0$ at 9L. In addition, decreasing the thickness of sample from 36L to 9L enhances the DOS, and leads to sharper pre-edge of the electronic DOS (Inset at the upper panel of Figure 3b) and increased hybridization of In 5$s$/5$p$ and Se 4$p$ orbitals (lower panel of Figure 3b). Note that the enhancement of DOS in the thinner InSe sample is fully consistent with Hicks-Dresselhaus's claim that reducing the width of quantum well increases DOS per unit volume in their pioneering work. Within the size range studied, we also observe the generally enhanced conduction-band charge accumulation in thinner samples due to stronger quantum confinement (Figure 3c-d).

The remarkable improvements of Seebeck coefficient in the thinner InSe samples results from the quantum confinement enhanced DOS, and may be expected to give rise to an increased thermoelectric power factor. We show the carrier density dependence of power factor at various temperatures (Figure 4a), which is defined as PF = $S^2\sigma_{3D}$, with $\sigma_{3D} = \sigma_{xx}/h_0$, $h_0$ being the thickness of sample. We observe that the PF value in the thinnest sample is increased by one order of magnitude with respect to the thick sample at similar carrier concentration. Particularly, the maximum PF= 0.18 mWm$^{-1}$K$^{-2}$ in the 29 nm thick InSe sample at 300 K can be increased to 2.3 mWm$^{-1}$K$^{-2}$ in the thinnest 7 nm sample at a common carrier density of 1.9×10$^{12}$ cm$^{-2}$ (Figure 4b). The PF value shows a significant increase by 13 times. According to the Hicks-Dresselhaus's theory[7], the decoupled thermopower and conductivity of in-plane transport can lead to an enhancement of the power factor in 2D material system when the dimensionality is reduced. This arises because of the different averages in the Boltzmann transport formulas for thermopower and conductivity. The enhancement of thermopower in the thin InSe sample is attributed to the enhancement of electronic density of states (DOS) at the Fermi level relative to bulk. In particular, by taking the common carrier density of $1.2 \times 10^{12}$ cm$^{-2}$, we found that the DOS value at the Fermi level increases from 0.16 states/eV/f.u. at 36L thickness to 0.30 states/eV/f.u. at 9L (Inset at the upper panel of Figure 3b). This PF in 7 nm thick In sample is comparable to the reported value for WSe$_2$ and monolayer graphene[16,23], and is competitive with some of the best bulk thermoelectric materials, e.g. within a factor of two of highly optimized Bi$_2$Te$_3$. For the much thinner InSe samples, the confinement becomes stronger and the power factor has a much larger enhancement, which is justified by the theoretical results (Figure S7 in the Supporting Information). Note that our thinnest InSe sample for the PF measurement is 7 nm. It is very challenging to obtain reliable experimental PF data for the samples thinner than 7 nm due to the very large contact resistance by larger band gap in the thinner sample[24,43,44]. Furthermore, the physical picture becomes complicated in the very thin 2D layers with the even stronger quantum confinement. In particular, the carrier mobility may start to show effects of increased scattering, *e.g.*, due to reduced dielectric screening[45]. In addition, other factors, such as the surface/interfacial scattering and the band bending effect in gated nanoflakes may also play non-negligible roles. However, we are not in that regime yet and it is not clear if and at what thickness these complications might become dominant.

Although the utilization of 2D quantum confinement has been proposed to engineer the DOS and the power factor in the quantum well thermoelectric materials, the critical condition associated with threshold thickness for realizing these benefits in 2D materials is somewhat ambiguous[9,10]. To clearly identify the critical condition for drastically enhancing power factor in the 2D materials at a specific temperature, the role of thermal de Broglie wavelength $\xi$ cannot be ignored in the thermoelectric theory as proposed before[11,12]. The thermal de Broglie wavelength is defined as $\xi = h/(2\pi k_B T m_e^*)^{1/2}$, where $h$ is Planck constant, $k_B$ is Boltzmann constant, $m_e^*$ is the effective mass of charge carrier (Table S1 in the Supporting Information). For degenerate semiconductors, the power factor PF can be written as a function of the ratio of the quantum confinement length $L$ over the thermal de Broglie wavelength $\xi$,

$$\text{PF} = \frac{4\pi^4 \mu K_B^2}{9qL^3} \left(\frac{L}{\xi}\right)^D \frac{(r+D/2)^2 \Gamma(\frac{5}{2})}{D\Gamma(\frac{5}{2}+r)\Gamma(\frac{D}{2})} \eta^{(r+\frac{D}{2}-2)}, \quad (1)$$

$q$ represents the electron charge, $\mu$ is the carrier mobility, $D$ is the dimensionality of sample, $r$ is the scattering exponent dependent on the dominant scattering mechanism of carriers, $\Gamma$ is the Gamma function, $\eta = (E_F - E_C)/K_B T$ is the reduced chemical potential (Fermi level ($E_F$) measured from conduction band edge ($E_C$) against $K_B T$). Note that $r = 0$ corresponds to the case of the acoustic phonon scattering. For 2D material, $D = 2$ and $L$ is largely determined by its thickness ($L \approx h_0$).

To quantitatively analyze the effect of quantum effects on PF, we present the PF value as a function of $h_0/\xi$ for three carrier densities in Figure 5. Note that the main goal of Eq. (1) used here is to demonstrate a strong $h_0/\xi$ dependence of the power factor PF. However, the Eq. (1) depends on the carrier mobility and it is inappropriate to use Eq. (1) with a constant mobility $\mu$ to analyze the experimental data (symbols) in Fig. 5. To exclude the effect of carrier mobility, we have normalized the theoretical PF over the constant mobility μ in the Eq. (1), with results shown in the inset of Fig. 5. In this case, the inset clearly shows a drastic enhancement of PF/μ when $h_0/\xi$ is less than 1, which is consistent with the experimental data shown in the main panel of Fig. 5. As seen in the Fig. 5, the change of the PF is little with respect to $h_0/\xi$ when $h_0/\xi$ is far larger than 1, the value of which reaches reported value (dashed line) for bulk InSe[26,27]. In stark contrast, The PF starts increasing significantly as the thickness of InSe sample is comparable to or smaller than the thermal de Broglie wavelength. For example, the PF at $h_0/\xi = 0.64$ is enhanced by 13 times compared to the PF of 0.18 mWm$^{-1}$K$^{-2}$ at $h_0/\xi = 2.6$ and $n = 1.9 \times 10^{12}$ cm$^{-2}$. As mentioned above, the power factor is enhanced by dimensional reduction due to a decoupling of conductivity and thermopower for in-plane transport. This is also corroborated by the recent experiments regarding the thermoelectric power factor of 2D electron gas[46,47]. The substantial enhancement of the PF within the regime of $h_0/\xi \leq 1$ indicates that the parameter $h_0/\xi$ plays an important role in optimizing the power factor and should be considered for designing the high-performance thermoelectric devices in the future. We note that the thermoelectric power factor enhancement upon the quantum confinement was recently reported in the 2D electron material $SrTi_{1-x}Nb_xO_3$ system[47], which was synthesized by pulsed laser deposition technique on insulating substrate. Compared to the non-layered $SrTi_{1-x}Nb_xO_3$ material, the 2D layered structure materials such as InSe herein may offer a more advantageous platform for precisely exploring the improved thermoelectric power factor in relation to the quantum confinement in the low-dimensional system, due to the high-quality samples with controllable, uniform and continuously-varying thickness and gate-tunable carrier density.

Bulk indium selenide compounds, such as $In_4Se_{3-\delta}$ and $InSe/In_4Se_3$ composites, have good thermoelectric performance, which is achieved by chemical doping and manipulation of the thermal conductivity[26,27]. Compared to these bulk compounds, better control over carrier density and quantum confinement in 2D InSe can be achieved to realize the higher power factor in 2D InSe materials at room temperature. Besides, we also note that the thermal conductivity of 2D InSe is low and perhaps may be further reduced by using strategies similar to bulk. Combining the enhanced power factor and the low thermal conductivity, 2D InSe-based thermoelectric devices are highly promising.

In summary, we investigate quantum confinement for enhancing the thermoelectric performance of 2D InSe thin films through experimental measurements and first-principles calculations. We observe a substantial enhancement of the Seebeck coefficient and the power factor by reducing well-controlled film thickness and modulating electron density by electrostatic gating. At room temperature, the maximum magnitude of Seebeck coefficient reaches ~570 μV/K in the 7 nm thick InSe film, much higher than the value reported of bulk samples. Based on the *ab initio* calculations and the Boltzmann transport theory, we ascribe the enhancement of the Seebeck coefficient to the increasingly strong quantum confinement in the thinner films. More importantly, we experimentally identify the key role of the ratio between the quantum confinement length and the thermal de Broglie wavelength, a previously-overlooked parameter, in optimizing the power factor of thermoelectric materials. Our results demonstrate a promising path for enhancing the power factor and thus thermoelectric performance of 2D materials and may pave the way towards utilizing 2D materials in practical thermoelectric devices.

## Methods

**Materials and devices**

InSe and hBN flakes were mechanically exfoliated onto PDMS and 285-nm-thick $SiO_2$ wafer substrates, respectively. The thicknesses of exfoliated flakes were confirmed by the Bruker Multimode 8 atomic force microscope (AFM). The InSe flakes were transferred onto the hBN substrate. Then InSe flakes on the hBN substrate were patterned into a thermoelectrics configuration by electron-beam lithography technique, followed by deposition of Ti/Au (5/50 nm) with standard electron beam evaporation. After the fabrication of thermoelectric devices, InSe thermoelectric devices were annealed in a pure argon atmosphere at $T = 270$ ℃ for 3 hours to remove polymer residues.

**Transport measurements**

All the electronic and thermoelectric transport measurements were carried out in an Oxford Instruments Teslatron$^{TM}$ CF cryostat. The four-terminal resistance of InSe devices and micro-thermometers were measured by the lock-in amplifier (Stanford Research 830) technique. Voltage was applied to the micro-heater by the multifunction data acquisition system (NI 6251) and the thermoelectric voltage was measured by a voltage amplifier (Stanford research systems model SR560).

**Electronic properties of InSe with various thickness**

Electronic structures of bulk and multiple-layers InSe were calculated based on density-functional theory (DFT) and the projector augmented wave method[48] with generalized gradient approximation of Perdew-Burke-Ernzerhof as implemented in the Vienna Ab initio Simulation Package[49,50]. The In $4d^{10}5s^25p^1$ and Se $5s^25p^4$ electron configurations were treated as valence electrons. We use a plane-wave basis kinetic energy cut-off of 520 eV and *k*-point mesh spacing 2π×0.03 Å$^{-1}$ or less for Brillouin zone integration. To properly consider the long-range van der Waals interaction that cannot be ignored in layered materials such as InSe, the vdW-optB86b functional is used[51]. We optimized all the structures by total energy minimization until the residual forces on all atoms were less than 0.1 meV Å$^{-1}$. To rectify the band-gap underestimation problem by normal DFT calculations, we used the Heyd-Scuseria-Ernzerhof (HSE06) hybrid-functional approach to correct band gaps[52]. The calculated band gap of 1.315 eV for bulk InSe based on a 35% nonlocal Fock exchange agrees with experiment value of 1.32 eV[53], which was applied to all the calculations presented in the paper.

**Boltzmann transport theory**

In the Boltzmann transport theory, Seebeck coefficient can be given as

$$S_{\alpha,\beta}(T,\mu) = \frac{\int (\varepsilon-\mu)\sigma_{\alpha\beta}(\varepsilon)[-\frac{\partial f(T,\varepsilon,\mu)}{\partial \varepsilon}]d\varepsilon}{qT\int \sigma_{\alpha\beta}(\varepsilon)[-\frac{\partial f(T,\varepsilon,\mu)}{\partial \varepsilon}]d\varepsilon}, \qquad (2)$$

where $\sigma_{\alpha\beta}(\varepsilon)$ is the energy dependent conductivity tensor, which depends on the energy dependent scattering time and the band velocities. $f(T,\varepsilon,\mu)$ is the Fermi distribution, and $\mu$ is the chemical potential. The $\varepsilon$ dependence of relaxation time is often taken as $\tau(\varepsilon) = \tau_0(\varepsilon/(k_BT))^r$, where $\tau_0$ is the relaxation time coefficient, $\varepsilon$ denotes the electron energy, $r$ is the scattering exponent dependent on the dominant scattering mechanism. In any case the Seebeck coefficient then depends on the energy dependence of the conductivity, including the band structure and scattering rate, i.e. σ(ε) ~ N(ε)<v$^2$(k,ε)>τ, where N(ε) is the DOS, <v$^2$(k,ε)> is the averaged square band velocity. For acoustic phonon scattering mechanism, which gives rise to $r = 0$ in the 2D systems this turns out to be the same as the constant scattering time approximation. Eigen energies at 5000 *k* points on a non-shifted mesh in the irreducible Brillouin zone were calculated to evaluate Seebeck coefficient through the BoltzTraP code[40]. This takes into account the effects of non-parabolicity and anisotropy of bands, multiple minima and multiple bands, etc. on thermoelectric transport.

## ASSOCIATED CONTENT

**Supporting Information**

The Supporting Information is available free of charge on the ACS Publications website at http://pubs.acs.org.

Seebeck coefficient measurement; Comparison between measured Seebeck coefficient and Mott relation; Electronic band structure and DOS of InSe with different thickness; Conductivity ($\sigma_{3D}$) of InSe with different thickness; The measured mobility in the samples with different thickness; Normalization of the power factor with the value of 29 nm; Calculated PF as a function of carrier density in much thinner InSe samples.

## AUTHOR INFORMATION

**Corresponding Authors**

* E-mail: miao@nju.edu.cn (F.M.).

* E-mail: lijun_zhang@jlu.edu.cn (L.Z.).

* E-mail: sjliang@nju.edu.cn (S.J.L.).

**Notes**

The authors declare no competing financial interests.

## ACKNOWLEDGMENTS

This work was supported in part by the National Key Basic Research Program of China (2015CB921600), the National Natural Science Foundation of China (61625402, 61574076, 11474147), the Natural Science Foundation of Jiangsu Province (BK20150055, BK20180330), and the Fundamental Research Funds for the Central Universities (020414380093, 020414380084) and the Collaborative Innovation Center of Advanced Microstructures. K.W. and T.T. acknowledge support from the Elemental Strategy Initiative conducted by the MEXT, Japan and the CREST (Grant No. JPMJCR15F3), JST. L.Z. acknowledges the support of the National Natural Science Foundation of China (Grant 11674121 and 61722403), the Recruitment Program of Global Youth Experts in China, and Program for JLU Science and Technology Innovative Research Team. Calculations were performed in part at High Performance Computing Center of Jilin University.

45, 4221-4226.

# Figure captions

**Figure 1.** Crystal structure and characterization of γ-InSe, thermoelectric device. (a) The crystal structure of InSe. (b) Raman spectrum of InSe for different thickness from 7 - 29 nm. (c) The schematic for thermoelectric measurement setup. (d) The optical microscopy image of a typical 7nm thick InSe device. The scale bar is 10 μm.

**Figure 2.** Electrical and thermoelectric transport. (a) Drain-source current $I_{ds}$ as a function of gate voltage $V_{bg}$ with a drain-source bias voltage $V_{ds} = 0.1$ V. (b) Four-terminal conductivity $\sigma_{xx}$ vs $V_{bg}$. Inset shows $T$-dependent mobility at two different carrier densities. (c) Seebeck coefficient $S$ versus carrier density $n$. The theoretical calculations (dashed lines) are based on ab initio calculations and Boltzmann transport equation. (The data shown in a-c is measured from a 10 nm device.) (d) Seebeck coefficient as a function of carrier density in different devices with thickness from 7 - 29 nm at room temperature. The dashed lines represent the theoretical results for corresponding thickness of sample.

**Figure 3.** Enhanced quantum confinement effect of InSe films with decreasing thickness. (a) Band structures of 9L and 36L thick InSe films. Red and blue colors represent projections onto constituting atomic species In and Se, respectively. The conduction band minimum (CBM) is set to energy zero and the effective mass values are indicated (in unit of $m_0$). (b) Total density of states (DOS) of each layer in 9L and 36L thick InSe films (upper panel), and the atomic orbital-decomposed DOS of 9L InSe film (lower panel). (c) Planar-averaged squared magnitude of wave functions for the CBM states of 9L and 36L thick InSe films, plotted along the direction perpendicular to layers. (d) Band decomposed charge density contour plots for the CBM states of 9L and 36L thick InSe films. The unit of charge density is e/Å$^3$.

**Figure 4.** Power factor of InSe films. (a) Power factor PF as a function of carrier density *n* in the 10 nm thick device. All the measurements are taken in the temperature region of 90 K - 300 K with the 30 K interval. (b) PF vs *n* in four devices with different thickness of sample at room temperature.

**Figure 5.** Enhancement of power factor based on the competition between the quantum confinement length ($h_0$) and the de Broglie wavelength ($\xi$). PF as a function of $h_0/\xi$ at three different carrier densities. The horizontal dashed line represents the experimentally reported PF values of bulk indium selenide compounds[26,27]. The inset represents the theoretical PF/$\mu$ versus $h_0/\xi$, which is consistent with the experimental data in the main panel.

# Figure 1

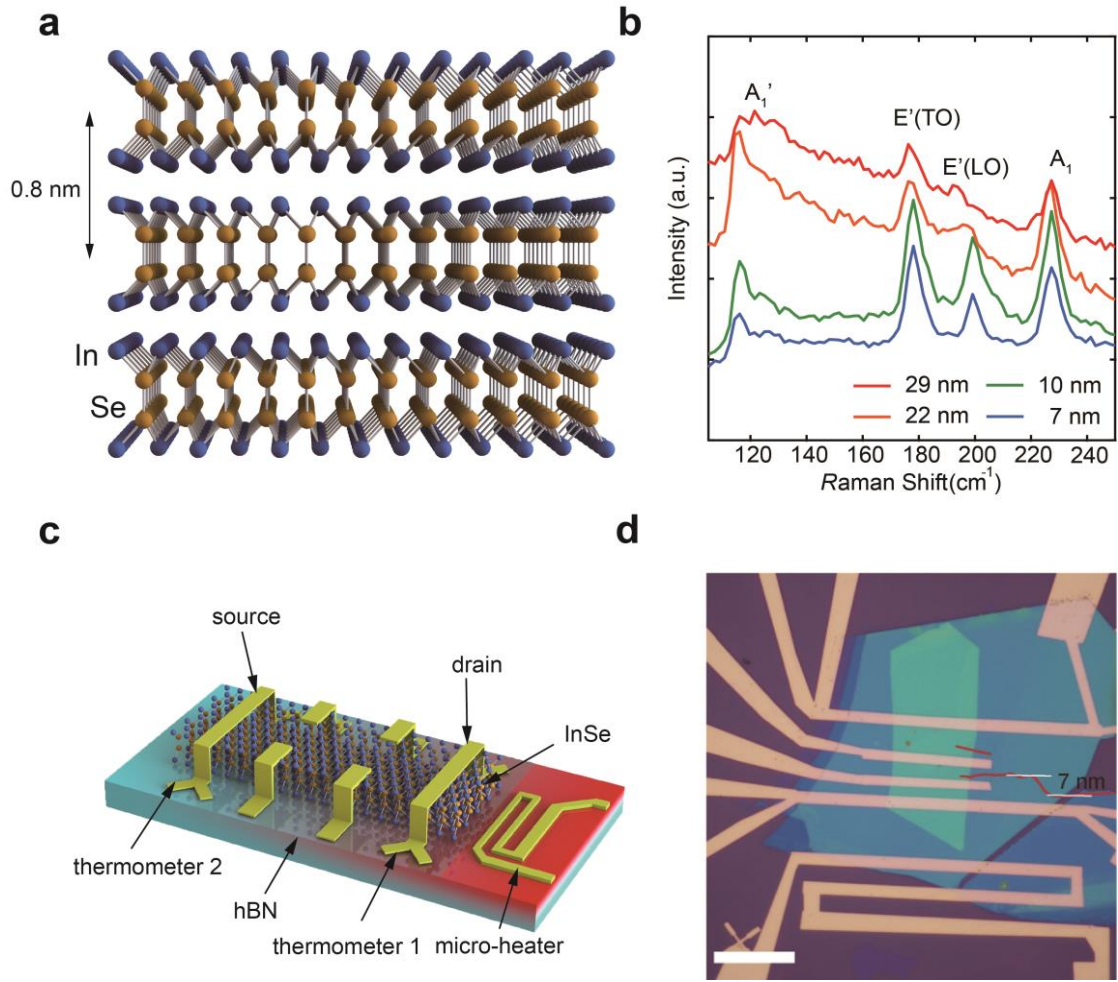

**Figure 2**

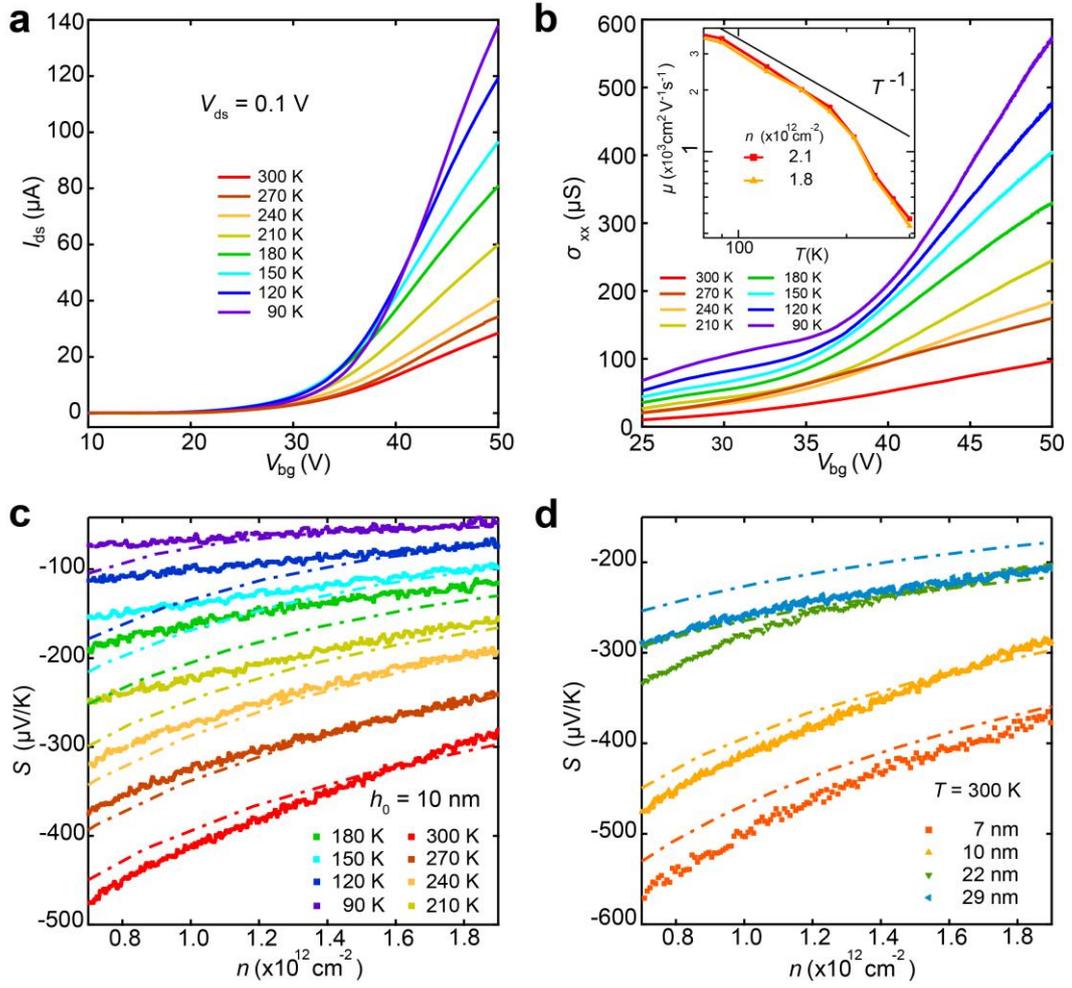

**Figure 3**

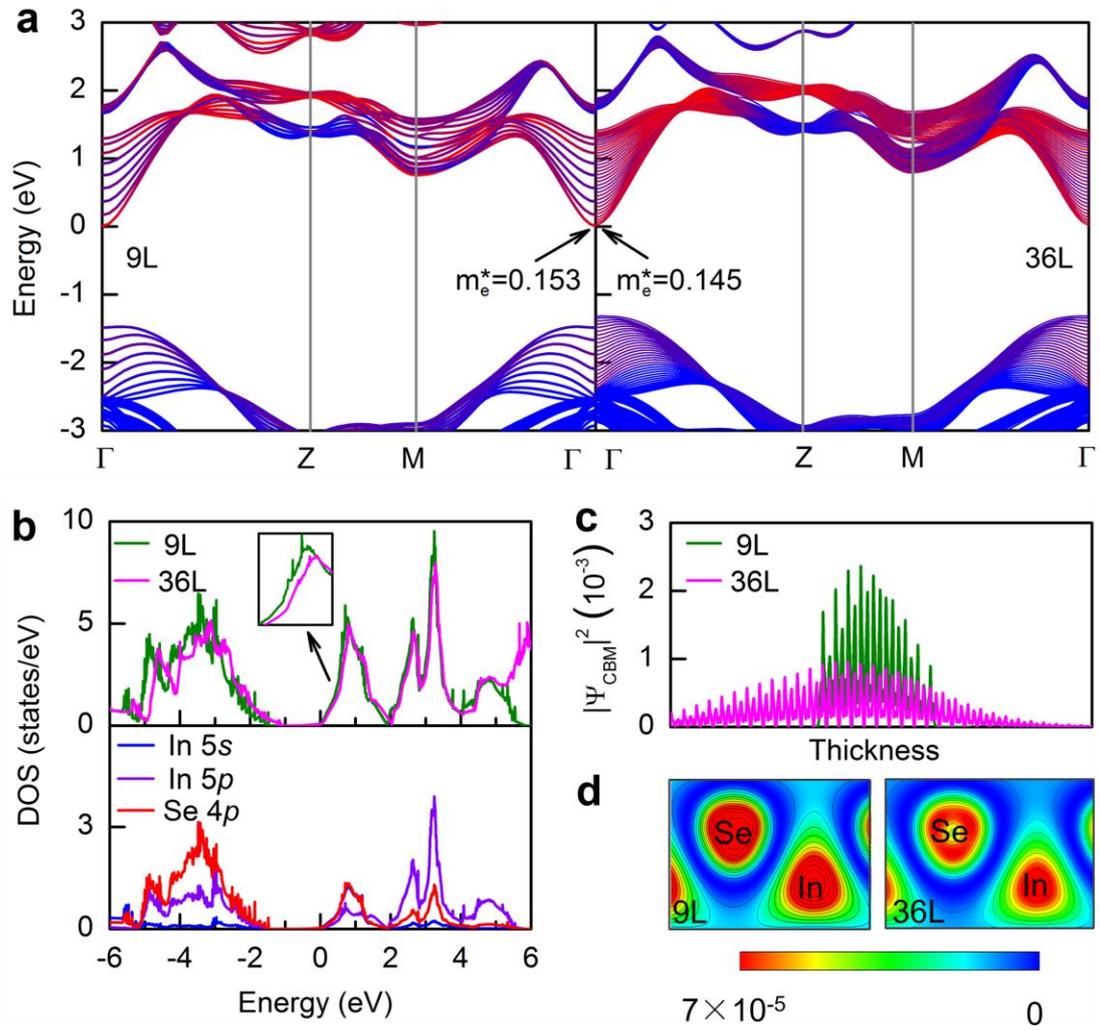

**Figure 4**

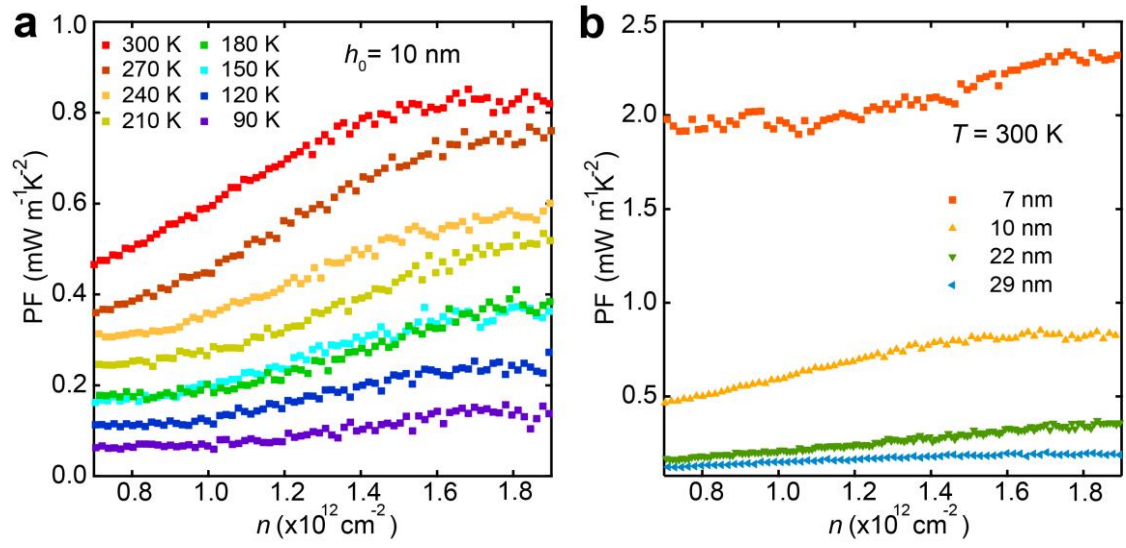

**Figure 5**

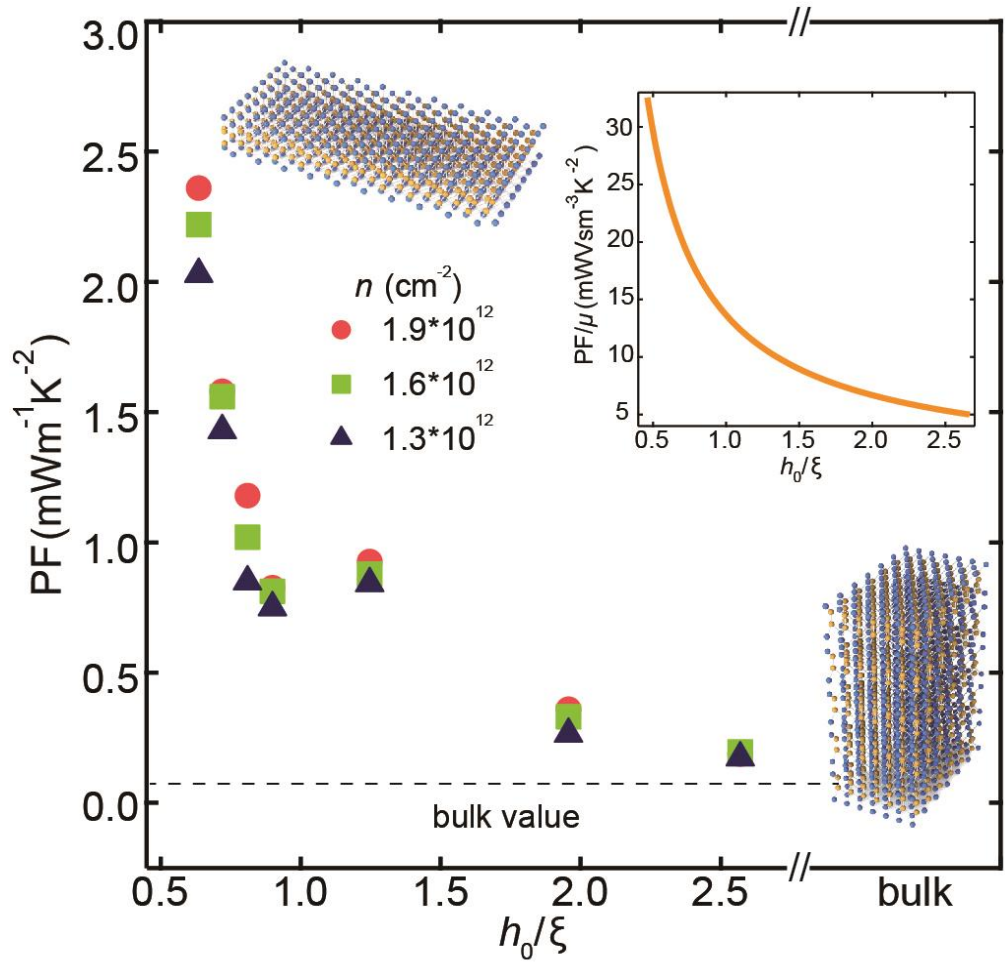